\documentclass[sigplan,10pt]{acmart}
\acmSubmissionID{1060}
\renewcommand\footnotetextcopyrightpermission[1]{}

\usepackage[]{hyperref}
\usepackage{xspace}
\usepackage{subcaption}
\usepackage{booktabs}
\usepackage{multirow}
\usepackage[linesnumbered,ruled,vlined,noend]{algorithm2e}
\usepackage{soul}
\newcommand{\struct}[1]{{$\langle${\color{red}{#1}}$\rangle$}}
\newcommand{\sys}[0]{{T-MAC}\xspace}
\newcommand{\jianyu}[1]{\textcolor{orange}{jianyu: #1}\xspace}

\newcommand{\revision}[1]{{{{#1}}}}
\newcommand{\remove}[1]{}

\newcommand{\subpara}[1]{{\it #1 \hspace{1pt}}}

\begin{document}

\acmYear{2025}\copyrightyear{2025}
\setcopyright{acmlicensed}
\acmConference[EuroSys '25]{Twentieth European Conference on Computer Systems}{March 30--April 3, 2025}{Rotterdam, Netherlands}
\acmBooktitle{Twentieth European Conference on Computer Systems (EuroSys '25), March 30--April 3, 2025, Rotterdam, Netherlands}
\acmDOI{10.1145/3689031.3696099}
\acmISBN{979-8-4007-1196-1/25/03}

\title{\sys{}: CPU Renaissance via Table Lookup for Low-Bit LLM Deployment on Edge}

\date{}

\settopmatter{authorsperrow=4}

\author{Jianyu Wei}
\authornote{Work done during the internship at Microsoft Research.}
\affiliation{%
 \institution{USTC / Microsoft Research}
 \country{China}
}
\email{noob@mail.ustc.edu.cn}

\author{Shijie Cao}
\authornote{Corresponding Author.}
\affiliation{%
 \institution{Microsoft Research}
 \country{}
}
\email{shijiecao@microsoft.com}

\author{Ting Cao}
\authornotemark[2]
\affiliation{%
 \institution{Microsoft Research}
 \country{}
}
\email{ting.cao@microsoft.com}

\author{Lingxiao Ma}
\affiliation{%
 \institution{Microsoft Research}
 \country{}
}
\email{lingxiao.ma@microsoft.com}

\author{Lei Wang}
\authornotemark[1]
\affiliation{%
 \institution{UCAS / Microsoft Research}
 \country{}
}
\email{t-leiwang@microsoft.com}

\author{Yanyong Zhang}
\affiliation{%
 \institution{USTC}
 \country{}
}
\email{yanyongz@ustc.edu.cn}

\author{Mao Yang}
\affiliation{%
 \institution{Microsoft Research}
 \country{}
}
\email{maoyang@microsoft.com}

\begin{abstract}

The deployment of Large Language Models (LLMs) on edge devices is increasingly important to enhance on-device intelligence. 
Weight quantization is crucial for reducing the memory footprint of LLMs on devices.
However, low-bit LLMs necessitate \textbf{mixed precision matrix multiplication (mpGEMM)} of low precision weights and high precision activations during inference.
Existing systems, lacking native support for mpGEMM, resort to dequantize weights for high precision computation.
Such an indirect way can lead to a significant inference overhead.

In this paper, we introduce \sys, an innovative lookup table(LUT)-based method designed for efficient low-bit LLM (i.e., weight-quantized LLM) inference on CPUs. 
\sys directly supports mpGEMM without dequantization, while simultaneously eliminating multiplications and reducing additions required. 
Specifically, \sys{} transforms the traditional data-type-centric multiplication to bit-wise table lookup, and enables a unified and scalable mpGEMM solution. 

Our LUT-based kernels scale linearly to the weight bit-width. Evaluated on low-bit Llama and BitNet models, \sys{} demonstrates up to 4$\times$ increase in throughput and 70\% reduction in energy consumption compared to \textit{llama.cpp}. \revision{For BitNet-b1.58-3B, \sys{} delivers a token generation throughput of 30 tokens/s with a single core and 71 tokens/s with eight cores on M2-Ultra, and 11 tokens/s on 
Raspberry Pi 5. }
\sys{} with LUT-based computing paradigm, paves the way for the practical deployment of low-bit LLMs on resource-constrained edge devices without compromising computational efficiency. \revision{The system is open-sourced at \url{https://github.com/microsoft/T-MAC}.}

\end{abstract}

\maketitle 



\pagestyle{empty}

\section{Introduction}
More and more large language models (LLMs) are deploying on client devices, such as smartphones, desktops, and robotics, to provide unprecedented intelligence services, real-time task response and user data protection. Typical examples are Phi-3-mini-4bit deployed on iPhone~\cite{phi3}, Llama-2-7B-4bit on Pixel 5 and Llama-2-13B-4bit on Apple M2 Ultra~\cite{llama.cpp}, as well as the recent Microsoft Copilot+PC~\cite{aipc} that collaboratively run the on-device LLM with the on-cloud LLMs. 

Low-bit weight quantization is a must-have for on-device LLM inference due to the hardware resource limitation. LLM inference quality is also robust to precision loss.  Beyond 4-bit, 3-bit, 2-bit and even 1-bit models are emerging~\cite{du2024bitdistiller,xu2024onebit,wang2023bitnet,chee2024quip}. By comparison, activation quantization cannot follow the trend due to outliers. The computing operands thus have asymmetric precision and bit width, such as W4A16, W2A16, or W1A8\footnote{W\# means the bit-width of weight and A\# means the bit-width of activation}. 

On the other hand, current commodity hardware still supports fixed bit-width and symmetric operands, and cannot support these various and mixed precision. Even some research papers support asymmetric bit-width operands. The bit-width is still fixed, such as W4A8~\cite{gope2020high}. Computing kernels have to convert/dequantize low-bit weight to match the activation, and compute on hardware. 

This conversion raises two obvious issues. \textit{(i)} Performance wise, the conversion cost offsets the gain from bit scaling down. Our evaluation  (ref Figure~\ref{fig:GEMV}) shows that scaling down bits from 4 bit to 1 bit even increases latency cost for most of the cases. \textit{(ii)} Development wise, the data layout and kernels need to be designed case by case for each mixed precision. For example, the data layouts, as well as the interleaving or swizzling methods for W3 and W2 are totally different. The kernels need to be redesigned to match the layout.    
Therefore, to deploy LLM on devices, a fundamental problem is how to directly and efficiently support mpGEMM (mixed precision General Matrix Multiplication) of low bit weights and high bit activations on devices. 

This paper aims a mpGEMM kernel design, which is independent of hardware data types and the bit width of the quantization algorithms, to achieve scalable speedup with bit scaling down. To realize that, our \textit{key concept} is that rather than the dominant data-type-centric calculation, we exploit the \textit{bit-wise} calculation of the standard algorithm of multiplication. That is, the multiplication of two numbers can be transformed into multiplying one number by each bit of the other number, then shifting and adding the partial products. 
The mpGEMM between activation and weight matrices is decomposed into a series (= the bit-width of weight) of mpGEMM between activation and a one-bit matrix, and then adds the partial results up. This method can thus support any bit-width combination of activation and weight. 

A promising method to implement bit-wise calculation is by table lookups~\cite{park2023lutgemm}. Since one bit can only represent two values, e.g., 1/-1, the bit patterns of a one-bit vector are limited. For example, if a one-bit matrix is partitioned into groups of four-element vector, the number of possible bit patterns (e.g., [1,1,1,-1] and [1,1,-1,-1]) for each group is only $2^4$. Given an activation, it can be first computed with all possible bit patterns and saved in tables. The mpGEMM of activation and one-bit matrix is then transformed to table lookup indexed by each bit pattern in weight, and addition to accumulate the looked-up results. The mpGEMM is reduced to table lookup+add operations and no multiplication. 

Though bit-wise LUT-based mpGEMM could reduce multiplications, how to efficiently implement it on real devices is challenging, since current hardware is highly tailored to multiplication. The \textit{outstanding difference} between bit-wise LUT and traditional mpGEMM is the two operands for LUT method are tables and index matrices, rather than activation and weight. The data format and layout of the two operands are thus critical to the inference speed. \textit{(i)} One challenge is that compared to the continuous data access for activation and weight, the access to tables are random. The residence of tables in fast on-chip memory will be particularly important for final inference performance. 
\textit{(ii)} However, the on-chip memory is limited, and the LUT method enlarges the on-chip memory usage compared to traditional mpGEMM. This is because the LUT needs to save the results of the activation vector multiplied with all possible bit patterns. This is exponentially more than the activation itself.    

\begin{figure}[t]
  \centering
  \includegraphics[width=0.95\linewidth]{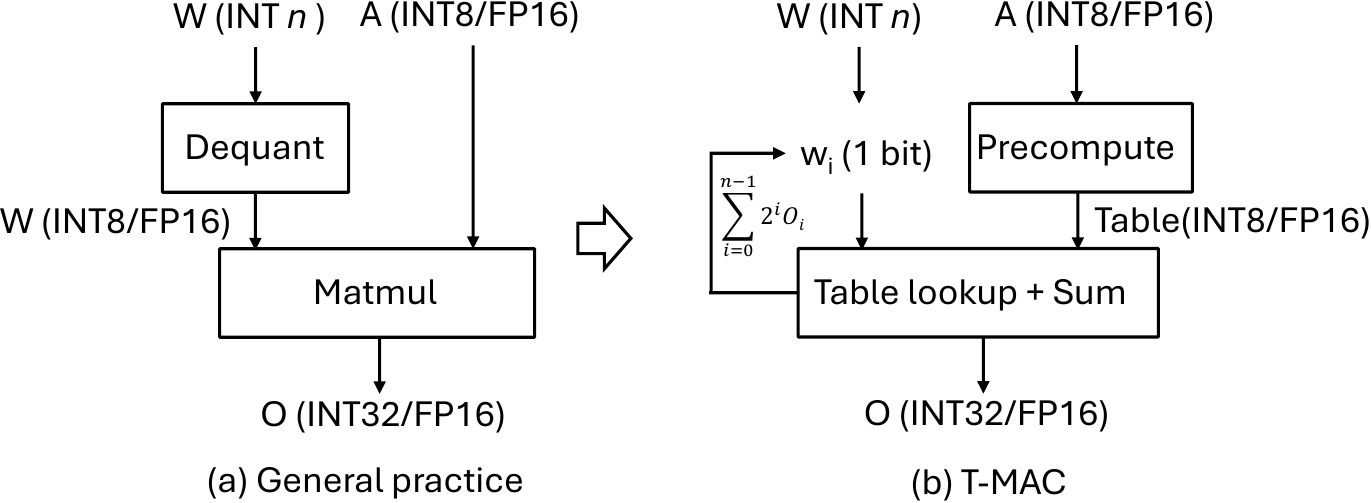}
  \vspace{-1em}
  \caption{\sys vs general practice for mpGEMM.}
  \vspace{-1.5em}
  \label{bittable}
\end{figure}

By solving the challenges, this paper proposes \sys mpGEMM kernel library. As shown in Figure~\ref{bittable}, it is based on the concept of bit-wise calculation and LUT implementation, and provide unified and scalable solution for any mixed bit-width of activation and weight. To alleviate the random-access cost of LUT, we propose techniques in both system and algorithm to enable the LUT to reside in the fastest on-chip memory and parallel lookup. On system-side, we propose the \textit{LUT-centric data layout}, to make sure a LUT in on-chip memory i.e., registers, through axis reordering and tiling to fully reuse each table as well as reduce the temporary results which competes with the on-chip memory. On algorithm-side, we propose \textit{table quantization} and \textit{mirror consolidation} to reduce the size of tables.


We implemented \sys{} on the pervasively available CPU processors of edge devices, even a Raspberry Pi. We find \sys{} makes the LLM inference speed on a CPU comparable or even higher to the GPU on the same device, mainly because \sys{} has no conversion cost and the total operation reduction by table lookup. \textit{This paper is thus the first to provide a practical solution for deploying LLMs on edge devices using the widely available CPUs, without relying on GPUs.}   
We evaluated \sys{} performance on typical edge devices, including Apple M2 Ultra, Jetson AGX Orin, Surface Book 3, and Raspberry Pi 5. The \sys{} kernel speedup can reach up to 6.6$\times$ and an average of 3.6$\times$ compared to the SOTA on the CPU by llama.cpp~\cite{llama.cpp}. The e2e LLM inference speedup achieves 2.8$\times$ speedup for Llama-2-7B-2bit model~\cite{gguf-models}. The inference performance can reach 11.1 tokens/s even on a Raspberry Pi for BitNet-b1.58-3B model~\cite{wang2023bitnet}. \sys{} is also energy efficient, reducing 60-70\% energy compared to llama.cpp.  

Our contributions can be summarized as follows:
\begin{itemize}
    \item \sys{} transforms the data-type-centric multiplication to \textit{bit-wise} table lookup for a unified and scalable mpGEMM design.  
    \item We propose both system and algorithm techniques to enable the table to reside on the fastest memory and parallel lookup.  
    \item We implement the \sys{} kernel library and e2e inference system to achieve significant LLM inference speedup and energy saving.
\end{itemize}

\section{Background and Motivation}

\subsection{LLM on Edge}
The advent of Large Language Models (LLMs) has revolutionized the field of natural language processing, opening new horizons for human-computer interaction and personalized assistant. The deployment of LLMs directly on edge devices, such as smartphones, desktop computers, and robotics, has emerged as a critical frontier in computing, promising to enhance these devices with unprecedented levels of intelligence and autonomy. 

\paragraph{LLM-on-edge benefits.}Deploying LLMs on edge devices brings several compelling benefits. On-device processing drastically reduces response latencies, which is critical for time-sensitive applications such as autonomous vehicles and interactive robotics. Furthermore, local data processing enhances user privacy by keeping sensitive information confined to the device, thereby reducing the risk of data leakage. Another significant advantage is the operational reliability that comes with network independence, allowing for consistent functionality regardless of network availability or stability.

\paragraph{LLM-on-edge challenges.}

The primary challenge in deploying LLMs on edge devices is the substantial memory requirement to accomodate these models. LLMs often encompass billions of parameters.
For example, LLAMA-2-7B with FP16 precision requires at least 14GB of memory to host the model. In contrast, edge devices typically have limited memory resources, which poses a stark limitation for on-device deployment. 


Beyond memory capacity, the computational and memory bandwidth demands of LLMs present a significant hurdle for edge deployment.
Edge devices often process data in a single-instance fashion, commonly with a batch size of one, to cater to a single user's real-time interactions. 
The inference process of the LLM can be divided into two stages: the prefill and decode. 
The prefill stage, with the self-attention mechanism applied across all input tokens, involves compute-intensive matrix-matrix multiplications.
However, once the key-value (KV) cache is generated, the decode stage becomes the bottleneck. In this phase, generating each subsequent token necessitates laoding and processing the entire model, which translates to memory-intensive matrix-vector multiplications. 

The third challenge is power or energy efficiency, which is particularly critical for battery-operated edge devices such as smartphones and robotics. These devices are designed for prolonged operation on finite energy reserves, making energy efficiency an essential consideration. 

\subsection{Low-Bit (Weight-Quantized) LLM}
The memory-intensive nature of LLM inference necessitates strategies to reduce the model's memory footprint without significantly compromising model performance. Weight quantization emerges as a key technique to achieve this balance.

Weight quantization involves reducing the precision of the model's parameters, effectively allowing the model to occupy less memory and potentially speed up computation by leveraging lower-precision arithmetic~\cite{dettmers2023case,frantar2022gptq,lin2023awq}. 
Nowadays, LLMs are increasingly being released with a 4-bit version specifically tailored for deployment on edge or other resource-constrained environments~\cite{team2023gemini,young2024yi}.
Recent research has pushed the boundaries even further, investigating the feasibility of 2-bit and 1-bit weight representations in LLMs~\cite{du2024bitdistiller,xu2024onebit,wang2023bitnet}. Fundamentally, the choice of bitwidth or precision in weight quantization represents a trade-off between computational efficiency and model accuracy.

\subsection{Deployment Challenges of Low-Bit LLM}

The adoption of low-bit LLMs has become a necessity for edge deployment. In fact, numerous LLM-on-Edge systems and implementations are actively employing low-bit techniques~\cite{llama.cpp,intel_neural_processor}.
However, deploying low-bit LLMs introduces unique computational challenges particularly in accommodating mixed-precision operations, which are not natively supported by most hardware architectures, and managing the diversity of bit widths and precision levels required by various deployment scenarios. Addressing these challenges is critical for full potential and seamless intergration of low-bit LLMs in edge computing.

\paragraph{Mixed-precison GEMM/GEMV}

The utilization of low-bit (weight-quantized) LLMs introduces a computational paradigm where low precision weights are combined with relatively higher precision activations. This necessitates a shift from standard matrix multiplication operations (i.e., GEMM, GEMV) to mixed-precision ones (i.e., mpGEMM, mpGEMV). 
However, current hardware architectures, including CPUs, GPUs, and NPUs, do not natively support mixed-precision operations. These architectures are traditionally optimized for standard operations where both operands share the same data type and precision level.  

In response to this limitation, existing systems implement an indirect approach by employing dequantization, which involves converting low-precision weights back to a higher precision to align with activation precision. This process enables the use of high-precision GEMM for low-bit LLM inference. For instance, systems like those used in the \textit{Intel Neural Compressor} and \textit{llama.cpp} rely on this dequantization-based technique. However, the efficacy of such methods is contingent upon the assumption that dequantization does not become a bottleneck and can be overlapped with memory loading. Since this indirect approach ultimately reverts to high-precision computation, it fails to fully capitalize on the benefits of low-bit weights, such as reduced memory usage and potentially faster computation.

\paragraph{Bit-width/precision diversity}
Beyond the challenge of facilitating mixed-precision operations, the diversity of bit-widths and precisions required by different deployment scenarios compounds the complexity. Depending on the task's difficulty and the specific requirements of the deployment environment, a variety of bit-widths may be selected to optimize performance. No single bit-width or precision setting can universally satisfy the diverse demands of all possible use cases. Consequently, this necessitates computational approaches capable of supporting a spectrum of low-bit widths, ensuring adaptability to the wide-ranging needs of edge computing tasks.  




\subsection{LUT-based Computation for Quantized Model}

A new trend in the computation of quantized models is the adoption of Lookup Table (LUT)-based methods.
For quantized Convolutional Neural Networks (CNNs), where both weights and activations are quantized to levels such as 4-bit, 2-bit, or 1-bit, 
DeepGEMM~\cite{ganji2023deepgemm} precomputes all
possible products of weights and activations, stores them in
a lookup table, and efficiently accesses them at inference
time to avoid costly multiply-accumulate operations.
Another example is for vector quantization, where activations are vector quantized, MADDNESS~\cite{blalock2021multiplying} and LUT-NN~\cite{tang2023lut} also transform GEMM computation to table lookups.

In the context of low-bit LLMs, or weight-only quantized LLMs, the LUT-based approach has been explored on GPUs~\cite{park2023lutgemm,maleki2023look}.
These methods leverage the GPU's shared memory or cache to store and access the lookup tables. However, despite the theoretical reduction in computational complexity, the practical kernel performance is worse than dequantization-based kernels in ~\cite{cutlass,bitblas}. 
For example, when tested with weight matrix shapes from real-world Llama-2 models on A100 GPU, the average latency of LUT-GEMM kernels ~\cite{park2023lutgemm} is 2.34$\times$, 1.87$\times$, and 1.75$\times$ longer than dequantization-based kernels in BitBLAS~\cite{bitblas} for 
\(W_{\text{INT4}}A_{\text{FP16}}\), \(W_{\text{INT2}}A_{\text{FP16}}\), and \(W_{\text{INT1}}A_{\text{FP16}}\)
mpGEMVs, respectively.
The suboptimal kernel performance is attributable to the constraints of the GPU's fixed architecture, which offers either inadequate storage capacity for the lookup tables or insufficiently rapid table access.  
In contrast, the exploration of LUT-based mixed-precision GEMM/GEMV on CPUs remains uncharted. Our work pioneers this investigation by examining the viability and performance implications of applying LUT-based methods to CPU-based inference of low-bit LLMs. 

\section{Design}
\label{sec:design}

\begin{figure}[t]
  \centering
  \begin{subfigure}[b]{\linewidth}
    \includegraphics[width=\linewidth]{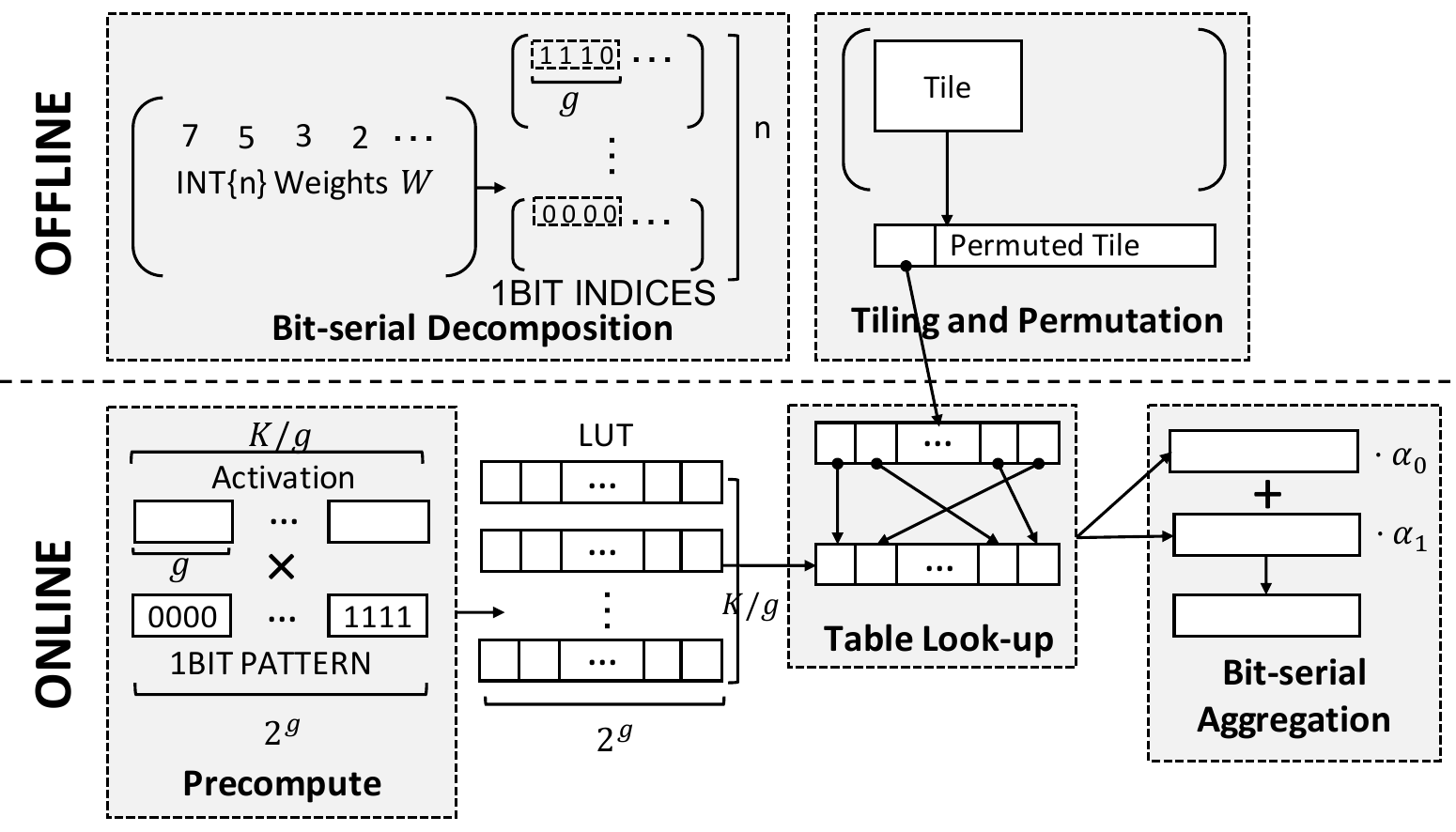}
  \end{subfigure}
  \vspace{-2.0em}
  \caption{\sys design overview.} 
  \label{fig:overview}
  \vspace{-2.6em}
\end{figure}

Current implementations for mixed-precision GEMM vary case by case. Each bit-width combination of activation and weight, such as W4A16 and W2A8, requires specific weight layout and computing kernels. For example, the layout for W3 could pack 2 bits and the other 1 bit in separate, and leverages different interleaving or swizzling methods for memory alignment or fast decoding. The corresponding computing kernel then needs to unpack this specific layout to a hardware-supported data type for execution.      

To provide a unified and scalable solution for mixed-precision GEMM, this paper transforms the dominant \textit{data-type-centric} computation to \textit{bit-wise} computation, based on the linear equivalent transformation in Eq.~\ref{eq:bit-wise}. For mixed-precision GEMM, $A$ and $W$ are the activation and weight matrix, respectively. $n$ is the bit-width of the weight. $W_i$ is each bit matrix of $W$. 
\vspace{-0.1in}
\begin{equation}
A\times W=A\times(\sum^{n-1}_{i=0}2^i W_i)=\sum^{n-1}_{i=0}2^i A\times W_i
\vspace{-0.1in}
\label{eq:bit-wise}
\end{equation}

In this way, the diverse weight layouts are reduced to a unified one-bit matrix layout. The diverse computing kernels are reduced to unified multiplication of the activation matrix and the one-bit matrix. Besides, bit-wise computation enables the linear scale-down of computation cost with the bit-width reduction.              

This paper exploits the LUT method to realize this bit-wise layout and multiplication (Sec.~\ref{subsec:lut_algo}), and proposes the LUT-centric data layout (Sec.~\ref{subsec:lut_data_layout}), as well as the table compression methods (Sec.~\ref{subsec:reduce_lut_storage}) to enable in-register table residence and the fastest parallel lookup.   

\subsection{\sys Algorithm}\label{subsec:lut_algo}

Figure~\ref{fig:overview} and  and Alg.~\ref{alg:sys} shows the \sys design. During the offline preparation stage (line 29 to 35), a $n$-bit weight matrix is decomposed into $n$ one-bit matrices. Since one bit can only represent two values, for a group with $g$ bits, the possible permutations are only $2^g$. \revision{During the online stage}, the permutations can be precomputed with each \remove{$g$-width vector of the activation}\revision{group of activation of shape $[1,g]$} and saved in a table. A $g$-bit group in the weight is thus an index to look up the table for the precomputed results. Therefore, a \textit{table} in \sys is defined to save the results of a $[1,g]\times[g,2^g]$ sub-matrix multiplication, and the table size is $[1,2^g]$. During the offline stage, a tile in the one-bit matrix will be saved continuously in memory to facilitate fast loading. Same as the tiling of normal matrix multiplication, a tiling here is also to improve data locality and cache utilization during LUT.

For the online stage, given the input activation of a GEMM, \sys traverses every  $[1,g]$ vector of the activation to multiply with the $[g,2^g]$ bit-pattern matrix and build up a table (line 16 to 27). During LUT, each index (i.e., group) of the one-bit weight matrix is used to look up the tables for partial results (line 6-9). The accumulation of the partial results will be the final GEMM results (line 12-14).

\revision{To illustrate with an example, given g=4, for an activation (A1,A2,A3,A4) of shape $[1,4]$ and 1-bit weights of shape $[4,M]$, the activation will be precomputed online into a LUT of shape $[1,16]$, containing elements from -A1-A2-A3-A4 to A1+A2+A3+A4. By grouping every 4 weights together, the weight vector of 0000 will lookup -A1-A2-A3-A4 and 0101 will lookup -A1+A2-A3+A4.}

\begin{figure}[t]
  \centering
  \includegraphics[width=\linewidth]{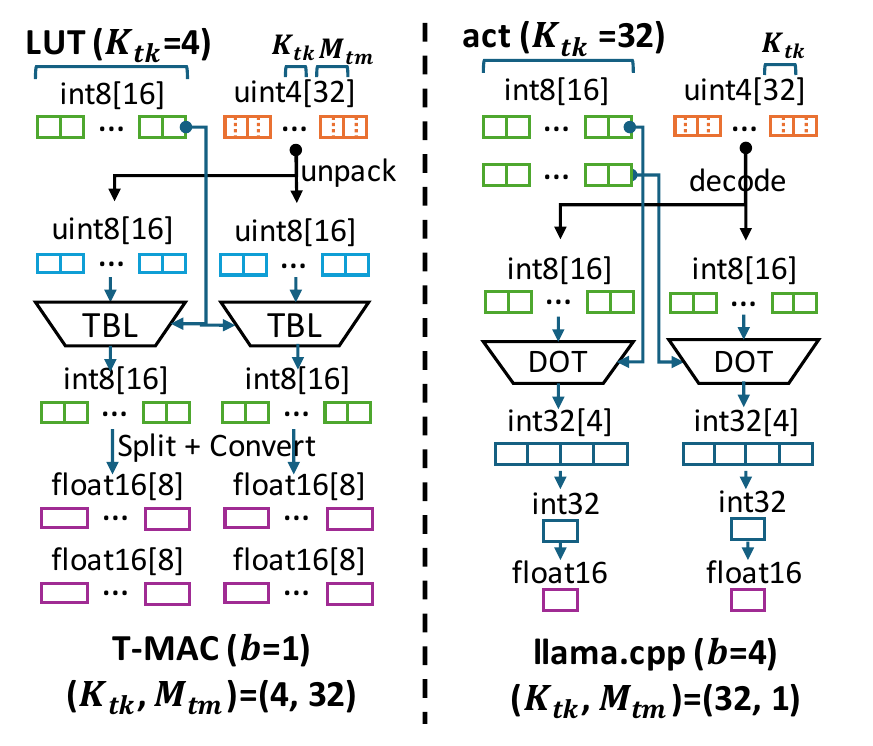}
  \vspace{-2em}
  \caption{Data flow of \sys vs general practice.}
  \vspace{-0.5em}
  \label{fig:dataflow}
\end{figure}

\paragraph{An example of LUT-based mpGEMM}
Figure~\ref{fig:dataflow} (left) shows an example of the \sys materialization on CPU. Taking the group size $g=4$, the tile size of the index matrix $W_i[K_{tk},M_{tm}]=[4,32]$, and the bit width as $b=4$. The right side is the general practice of mpGEMM implementation using llama.cpp. For \sys on the left side, 
the 32 uint4 indices are first unpacked into uint8 bytes (blue) to ensure compatibility with the hardware data type and instructions. Subsequently, the uint8 indices are utilized to look up the table. The results from the look-up are then split and converted to a higher precision for multiplying with the quantization scales of the low-bit LLM model.

By comparison, the general practice designs specific computing kernels for this 4-bit model. It first directly decodes the 4-bit weight to int8 to align with hardware data type, and then conducts int8 dot-product for the activation and weight vectors. Similarly, the result will be covered into FP16 for multiplying with quantization scales. The cache tiling for llama.cpp is $W[K_{tk},M_{tm}]=[32,1]$. Sec.~\ref{subsec:lut_data_layout} will explain the different tiling rationale of \sys from current practice.



\paragraph{Challenges of LUT implementation} From the algorithm and example, it can be seen that LUT-based mpGEMM exposes the following challenges. \textit{(i) Random data access.} Compared to the continuous data access of current practice, the tables are randomly accessed given the indices. It is necessary to host tables in the fast on-chip memory to reduce the access cost. \textit{(ii) Enlarged on-chip memory usage.} However, LUT requires more on-chip memory compared to current practice. The table size grows exponentially with the group size $g$. For instance, when $g=4$, the LUT is four times larger than the original activation. Besides, in contrast to the traditional GEMM implementation that yields \textit{scalar} outputs for each basic block, LUT method results in \textit{vector} outputs (as shown in Figure~\ref{fig:dataflow}), which requires more on-chip memory to store the temporary results. Considering the example in Figure~\ref{fig:dataflow} again, the LUT method uses 144 8-bit registers and llama.cpp uses 104 8-bit registers.      

To address the above issues, we propose two main techniques: (a) \textit{LUT-centric data layout} to accommodate intermediates and LUT into memory with higher bandwidth, and (b) \textit{reduced LUT storage} to decrease the LUT size and limit the number of look-up operations.

\begin{algorithm}
    \SetAlgoLined
    \DontPrintSemicolon
    \SetKwInOut{Input}{input}
    \SetKwInOut{Output}{output}
    \Input{Activation $A$ of shape $N$, $K$,\\weights $W$of shape $M$, $K$}
    \Output{Result matrix $R$ of shape $N$, $M$}
    let $b$ be the number of bits in weights\;
    $W_1$, ..., $W_b$ $\leftarrow$ PreprocessWeights($W$, $M$, $K$)\;
    LUT $\leftarrow$ Precompute($A$, $N$, $K$)\;
    let $\alpha_i$ ($i$ $\leftarrow$ 1 to $b$) be the multiplier of bit serial\;
    let $\beta$ be the bias of bit-serial\;
    \For{$i$ $\leftarrow$ 1 to $b$}{
        \For{$n$, $m$ $\leftarrow$ 1 to $N$, $M$}{
            $R_i$[$n$, $m$] = $\sum_{k \leftarrow 1}^{K}$ Look-up($LUT$, $W_i$, $n$, $m$, $k$)\;
        }
    }
    \;
    let $B$ be a matrix of shape $M$, $K$ with all elements $\beta$\;
    $R_\beta$ $\leftarrow$ $A \cdot B^T$\;
    $R$ $\leftarrow$ $\sum_{i \leftarrow 1}^{b}\alpha_i R_i + R_\beta$\;
    \;
    \SetKwProg{Fn}{Function}{:}{}
    \SetKwFunction{FMain}{Precompute}
    \Fn{\FMain{$A$, $N$, $K$}}{
        let $g$ be the group size of LUT table\;
        \For{$n$, $k$ $\leftarrow$ 1 to $N$, $K$}{
            \For{$i$, $j$ $\leftarrow$ 1 to $2^g$, $g$}{
                \uIf{$i$ \& (1 << $j$)}{
                    LUT[$n$, $k/g$, $i$] += $A$[$n$, $k$]\;
                } \Else{
                    LUT[$n$, $k/g$, $i$] -= $A$[$n$, $k$]\;
                }
            }
        }
        \Return LUT\;
    }
    \;
    \SetKwProg{Fn}{Function}{:}{}
    \SetKwFunction{FMain}{PreprocessWeights}
    \Fn{\FMain{$W$, $M$, $K$}}{
        \For{$i$ $\leftarrow$ 0 to $b$}{
            \For{$m$, $k$ $\leftarrow$ 1 to $M$, $K$}{
                $W_i$[$m$, $k/g$] += (W[$m$, $k$] >> i) << (k \% g)\;
            }
        }
        \Return $W_1$, ..., $W_b$\;
    }
    \caption{\sys GEMM}
    \label{alg:sys}
\end{algorithm}

\vspace{-1.0em}


\subsection{LUT-Centric Data Layout}\label{subsec:lut_data_layout}

As described in \S\ref{subsec:lut_algo}, the LUT-based method for low-bit GEMM requires more memory to store the lookup table and the intermediate results.
On the other hand, the table look-ups usually lead to memory access inefficiency due to the random memory accesses.
To resolve the inefficiency on memory storage and accesses, we design a LUT-centric data layout for the LUT-based low-bit GEMM. Specifically, it stores the lookup table on on-chip memory like registers to accelerate table accesses, and designs axis reordering and data tiling to enhance data reuses for reducing memory consumption.
Furthermore, to improve the efficiency, we design two data layout optimizations, i.e., weight permutation to align with memory transactions, and weight interleaving for optimizing weight unpack.

\paragraph{Put lookup table on on-chip memory.}
The LUT-based GEMM described in \S\ref{subsec:lut_algo} requires fetching the result from the pre-computed table, which should access the lookup table randomly. To accelerate these table look-ups, we put the lookup table on registers, and leverage the hardware-specific instructions (e.g., TBL on ARM CPUs and PSHUF on x86 CPUs) to do table look-ups. The details of optimizing table look-ups with hardware-specific instructions will be described in \S\ref{sec:impl}.
However, putting the lookup table on registers further increases the memory pressure on the on-chip memory, which may result in dramatic performance drop due to memory spilling. To fully explore the potential of on-chip memory table look-ups, we design axis reordering and tiling to enhance data reuses to reduce the on-chip memory pressure.

\paragraph{Axis reordering.}
For GEMM $C[N,M]=A[N,K] \times W[M,K]$, it is natural to loop among the spatial axes $N$ and $M$, and then the temporal axis $K$. However, the LUT-based GEMM needs to build the table among the $K$ axis, resulting in extreme large table storage when looping spatial axes then temporal axis following the traditional GEMM, i.e., a lookup table for the whole $A[N,K]$
But if we swap the axis order from spatial first to temporal first, it will only maintain a small lookup table $[1,K]$.
Therefore, \sys{} reorders the axes access to temporal axis $K$ first and then spatial axes $N$ and $M$.

\paragraph{Tiling.}
Tiling is a common technique in optimizing data locality of GEMM and reducing the memory requirements by reusing data on on-chip memory. Assume the GEMM $C[N,M]=A[N,K] \times W[M,K]$ is processed with tile $A[N_{tn},K_{tk}]$ and $W[M_{tm},K_{tk}]$, processing a tile requires $N_{tn}*K_{tk}+M_{tm}*K_{tk}$ data loading from DRAM to processor's on-chip memory instead of $N_{tn}*M_{tm}*K_{tk}$ data loading. In traditional GEMM, the tile size $N_{tn}$ and $M_{tm}$ have equal effects on efficiency while $K_{tk}$ does not affect the data reusing and is set to align the memory transaction.

However, in LUT-based GEMM, the activation $A[N,K]$ should be processed to build the lookup table, while the weight $W[M,K]$ can share the same pre-computed lookup table. That is to say, a larger tile size $M_{tm}$ on $M$ can lead to better lookup table reusing. \sys{} will carefully consider the tiling configurations $N_{tn}$, $M_{tm}$ and $K_{tk}$ to achieve better data reusing.

\begin{figure}[t]
  \centering
  \includegraphics[width=0.95\linewidth]{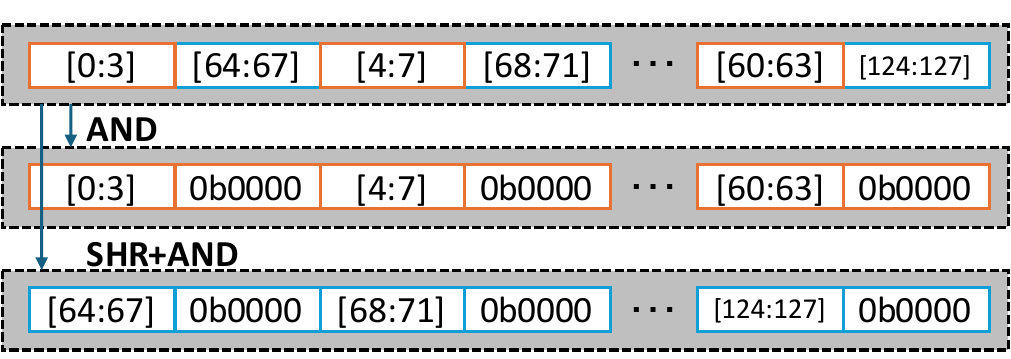}
  \vspace{-1em}
  \caption{Interleave weights for fast unpacking.}
  \vspace{-0.5em}
  \label{fig:interleave_weight}
\end{figure}

\paragraph{Layout optimizations.}
Beyond optimizing the on-chip table lookups with axis reordering and tiling, we design two data layout optimizations, i.e., weight permutation and interleaving, to further improve the efficiency.

\subpara{Weight permutation for sequential memory access.}
DRAM requires sequential accesses to achieve higher bandwidth utilization, while accessing tiles in the tiling-based GEMM introduces random accesses because tiles of the input matrices are not sequentially stored.
To solve this problem, \sys{} designs a weight permutation method that permutes the weight matrix to align the weight load with the memory transaction.
After the scheduling of a GEMM is determined, \sys{} will permute the input matrices to let tiles stored sequentially instead of the whole matrices. Specifically, \sys{} flats the elements in a tile sequentially and then concatenates the flatten tiles according to the tile accessing order.
Note that the weight matrix will not be modified during LLM inference, so this permutation can be done in offline. This offline permutation does not introduce cost in inference.

\subpara{Weight interleaving for fast unpacking.}
As described in \S\ref{subsec:lut_algo}, the weight matrix is stored in packed format in memory and should be unpacked during computation.
However, due to the commonly-used little-endian in modern CPUs, bytes in an integer are stored in a backward order. Therefore, using the integer instructions to unpack the weights requires additional reordering to return the correct unpacked weights.
As the weight matrix will not be modified during LLM inference, \sys{} can interleave the packed weights to eliminate this reordering. Figure~\ref{fig:interleave_weight} shows an example of this interleaving that unpacking the interleaved weights can directly produce the required weights sequentially.

\subsection{Reduce LUT Storage}
\label{subsec:reduce_lut_storage}

\begin{figure}[t]
    \centering
    \includegraphics[width=\linewidth]{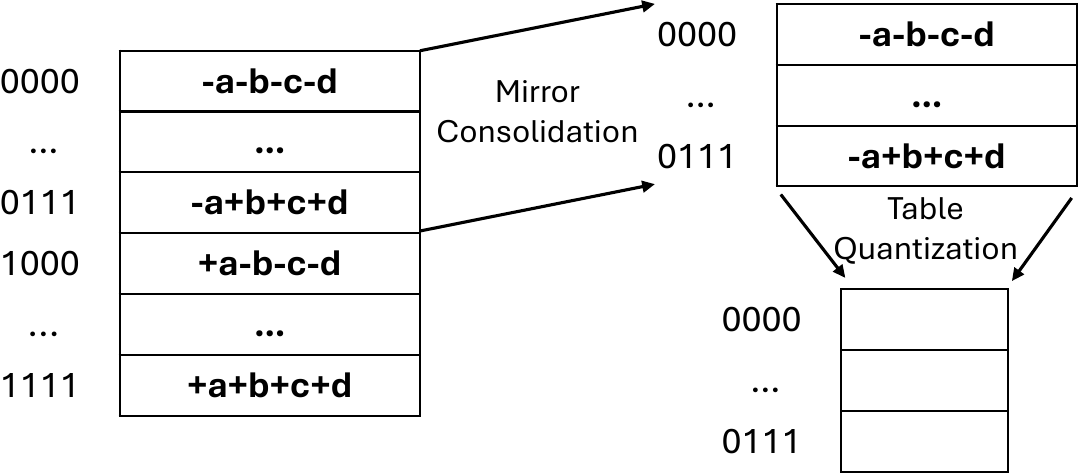}
    \caption{Reduce LUT storage with mirror consolidation and table quantization. Mirror consolidation halves the table length. Table quantization reduces the table width.}
    \vspace{-0.5em}
    \label{fig:table_reduction}
\end{figure}

In the realm of LUT-based method for low-bit LLM inference, the size of the lookup table is a crucial factor that impacts both the storage requirements and the access latency, especially when optimizing table look-ups with on-chip memory in \S\ref{subsec:lut_data_layout}.
A larger table size not only demands more memory space, but also leads to slower table access speed.
To address this challenge, we introduce two optimization strategies: \textit{mirror consolidation} and \textit{table quantization}. 
As illustrated in Figure~\ref{fig:table_reduction}, mirror consolidation exploits the symmetrical properties of table values to halve the length of the table, while table quantization applies quantization techniques to the table values themselves to reduce the width of the table. Combined, these methods enable a significant reduction in the storage footprint of the lookup table (up to a quarter of its original size) without accuracy loss in the LLM inference. 

\paragraph{Mirror Consolidation.}
The inherent symmetrical properties of table values in the context of lookup tables for LLM inference present a unique opportunity for optimization. Each positive value within the table is naturally paired with its negative counterpart, reflecting a mirror image across the zero value. Leveraging this symmetry, our proposed \textit{Mirror Consolidation} technique capitalizes on the fact that only half of the table values need to be explicitly stored. The remaining half can be rapidly reconstructed by simply negating the stored values. This table compression method is lossless, preserving the model's inference accuracy entirely. Furthermore, it proves to be highly efficient, accelerating the precomputation of the lookup table, reducing the required storage, and speeding up table accesses.

\paragraph{Table Quantization.}
Table quantization operates on a principle analogous to weight and activation quantization, aiming to reduce the precision of table values for improved computational efficiency. For instance, values initially represented in 16-bit floating-point (fp16) within a lookup table can be quantized to 8-bit integer (int8) with a scaling factor.
The impact of table quantization on model accuracy is negligible. Contrary to activation quantization, which has been challenging in maintaining model accuracy due to its necessity for coarse granularity and static quantization to ensure fast computation, our approach embraces finer granularity (quantizing 8 values for k=4) and dynamic quantization to minimize accuracy degradation. The findings presented in \S\ref{error_analysis} demonstrate that our table quantization technique has an imperceptible effect on the overall model accuracy. 
In terms of efficiency, table quantization significantly reduces lookup table storage requirements and accelerates the lookup process. 

\section{Implementation}\label{sec:impl}

\begin{table}[t]
    \centering
    \resizebox{\linewidth}{!}{
    \begin{tabular}{c c c}
        \toprule
        Instruction Set & Look-up & Fast Aggregation \\
        \midrule
        NEON & vqtbl1q\_u8 & vrhaddq\_u8 \\
        AVX2 & \_mm256\_shuffle\_epi8 & \_mm256\_avg\_epu8 \\
        \bottomrule
    \end{tabular}
    }
    \caption{Hardware Intrinsics for Look-up and Aggregation}
    \label{tab:intrinsics}
    \vspace{-2em}
\end{table}

\paragraph{Code generation through TVM}
We employ TVM~\cite{tvm} + LLVM~\cite{llvm} for code generation. This allows us to generate optimal code for GEMM of varying shapes and for different hardware, and implement common optimizations such as loop unrolling, vectorization, and constant folding. We utilize TVM Tensorize to embed hardware intrinsics into the code. AutoTVM~\cite{autotvm} is used to automatically fine-tune the generated code for different hardware targets.

\paragraph{API and integration} We provide a consistent API for both C++ and Python. The GEMM functions are encapsulated into TVM PackedFunc, and the tensors can be transferred through the DLPack~\cite{dlpack} tensor structure. This facilitates easy interoperability with other frameworks like PyTorch, Numpy, and etc. Additionally, we provide an additional wrapper for C++, where tensors can be passed through raw pointers and TVM runtime dependency is eliminated. This offers a more lightweight solution for integration into other C++ projects.

\paragraph{Parallelism} We utilize the TVM runtime threadpool to dynamically assign tasks to CPUs. However, when integrating \sys into llama.cpp, we notice an obvious conflict between the llama.cpp threadpool and the TVM threadpool. This conflict arises as threads from different threadpools compete for CPU resources, leading to a significant performance degradation. To resolve this issue, we generate C++ code using TVM, rather than directly creating library files, and then postprocess the code to remove the dependency on the TVM runtime and threadpool. The generated function will only execute computations of a single threadblock, and then we assign these threadblocks to different threads in the llama.cpp threadpool. This approach achieves better performance and compatibility with llama.cpp. The portable C++ code also offers an option for cross-platform deployment.

\paragraph{Efficiant table look-up by TBL/PSHUF} After loading the table into registers, we can utilize hardware intrinsics provided by ARM NEON/INTEL AVX2. Both NEON/AVX2 offer 8-bit look-up instructions. The bit width of ARM NEON is 128, which can precisely accommodate the entire table of $g=4$. INTEL AVX2 has a bit width of 256, but the lower and upper halves are in separate 128-bit lanes. Therefore, we duplicate the table to fill the 256-bit LUT register and look up 32 different int8 weight indices with a single instruction. If the table's data type is float16, since NEON/AVX2 do not support 16-bit LUT, we split float16 into two 8-bit LUTs, one for the lower part and the other for the higher part. We can look up the lower and higher parts with two instructions and then recombine them into float16.

\revision{
\paragraph{Determine the size of on-chip LUT} The number of on-chip LUTs is tuned for each hardware, to make sure on-chip memory can be fully utilized and LUTs won’t be swapped out for the tile. Larger on-chip memory allows more LUTs to reside, enabling more intermediate results to be aggregated before writing back. Register spill caused by excessive on-chip LUTs can lead to overhead.

$g$ is also determined by the on-chip memory size and instruction throughput. LUT for $g=4$ exactly fits into one register for ARM.TBL/AVX2.PSHUF. A larger $g$, such as $g=5$, requires two registers and slower ARM.TBL2/AVX512.PSHUF. 
}

\paragraph{Fast 8-bit aggregation} Besides table look-up, aggregation is another significant computational overhead. To speed up the aggregation, we initially aggregate the look-up results in low bit and later convert the aggregation sum to a higher precision like float16 with no accuracy loss. Moreover, we can implement fast 8-bit aggregation~\cite{blalock2021multiplying} if the table is quantized to int8. Normally, the int8 values should be converted to int16 to avoid overflow. However, int16 instructions have half the throughput of int8 aggregation. As an alternative, we can use \textit{avg/rhadd} instructions to compute the average and minimize accuracy loss by subtracting the probabilistic bias from the final value. Notably, fast 8-bit aggregation could result in nonnegligible accuracy loss.
In Table~\ref{tab:intrinsics}, we list the hardware intrinsics for look-up and aggregation on different CPU architectures.

\paragraph{Bit-serial linear transformation.}
In the decomposition mentioned in \S\ref{subsec:lut_algo}, we utilize the original values of $v_i$, i.e., 0 and 1. However, we can introduce a linear transformation to these values. Denote this linear transformation as $f(v_i)$ and the transformed values as $f(0) = s_0$ and $f(1) = s_1$.

To speed up precomputation and reduce quantization error, the values of $s_0$ and $s_1$ need to be chosen with care. To circumvent float-multiply instructions, we select them from the set [-1, 0, 1]. To reduce quantization error, we strive to minimize the difference between the largest and smallest entries of the lookup table (LUT). From empirical studies, we have found that $s_0 = -1$ and $s_1 = 1$ are optimal choices.

By setting the values of $s_0$ and $s_1$, we define the linear transformation $f$ and the decomposition of $W$ should be adjusted as follows:
\vspace{-0.1in}
\begin{align*}
    f(v_i) &= \alpha_i' v_i + \beta_i', v_i = \alpha_i f(v_i) + \beta_i, \\[0em]
    &\text{where}~\alpha_i = \frac{1}{\alpha_i'}, \beta = -\frac{\beta_i'}{\alpha_i'} \\[0em]
    W &= \sum_{i=0}^{b-1} \alpha_i 2^i W_i' + B, \\[0em]
    &\text{where}~W_i' = f(W_i), B = J \cdot \sum_{i=0}^{b-1} \beta_i 2^i \\[0em]
    &\text{and J is a matrix of ones}
\end{align*}
\vspace{-0.3in}

\paragraph{Register swizzling for efficient LUT precomputation}
As demonstrated in \S\ref{subsec:lut_algo}, we opt for subtraction/addition instructions over multiplication instructions to achieve better throughput. For the LUT of shape $(N, K/g, 2^g)$, the subtraction/addition can be vectorized along the $K/g$ axis. For instance,
\vspace{-0.05in}
\begin{align*}
LUT[0, 0:8, 0] = &-A[0, 0:32:4] - A[0, 1:32:4]\\
&- A[0, 2:32:4] - A[0, 3:32:4]
\vspace{-0.05in}
\end{align*}

The indexing into $A$ is not contiguous. By employing \textit{LD4} in NEON and \textit{vgatherdps} in AVX2, we can efficiently load non-contiguous data. However, when writing the non-contiguous LUT back into memory, extracting specific bytes from a SIMD register and writing to memory is highly inefficient for AVX2. To address this issue, we use register swizzling to rearrange the LUT so that it can be written back to memory contiguously. Initially, we use \textit{vpblendvb} to blend 8-bit values from different registers into one register, followed by \textit{vpermd} to swizzle the 32-bit values of the 256-bit, and then \textit{vpshufb} to further shuffle the 8-bit values into the correct order. After swizzling, the LUT can be written back to memory in a contiguous manner.

\section{Evaluation}

\begin{table}[t]
\centering
\resizebox{\linewidth}{!}{
\begin{tabular}{c c c c}
\toprule
\multirow{2}{*}{\textbf{Device}} & \multirow{2}{*}{\textbf{Processor}} & \textbf{Performance} & \textbf{Max. Memory} \\
 & & \textbf{Cores} & \textbf{Bandwidth} (GB/s) \\
\midrule
M2-Ultra & Apple M2-Ultra & 16 & 819.2 \\
Raspberry Pi 5 & ARM Cortex-A76 & 4 & 17.1 \\
Jetson AGX Orin & ARM Cortex-A78AE & 12 & 204.8 \\
Surface Book 3 & Intel Core i5-1035G7 & 4 & 58.2 \\
\bottomrule
\end{tabular}
}
\caption{Hardware device specification.}
\vspace{-2.0em}
\label{tab:spec}
\end{table}

We evaluate \sys with real-world, low-bit LLMs, specifically Llama and BitNet, across four distinct edge devices. Our benchmarking efforts are aimed at a direct comparison with the existing state-of-the-art \textit{llama.cpp} implementation. We summarize our key findings as follows:
\begin{itemize}
    \item The mpGEMV and mpGEMM kernels of \sys show a marked performance gain, significantly outperforming the state-of-the-art dequantization-based kernels.
    \item \sys enables an end-to-end model inference throughput improvement of 2-4x, while concurrently reducing the energy consumption by 60\%-70\% relative to the original \textit{llama.cpp} implementation.
    \item Remarkably, in many cases, \sys not only matches but even exceeds GPU performance, indicating a new milestone for LLM deployment efficiency on devices.
\end{itemize}

\subsection{Evaluation Setup}

\begin{figure*}[t]
  \centering
  \begin{subfigure}[b]{\linewidth}
    \includegraphics[width=\linewidth]{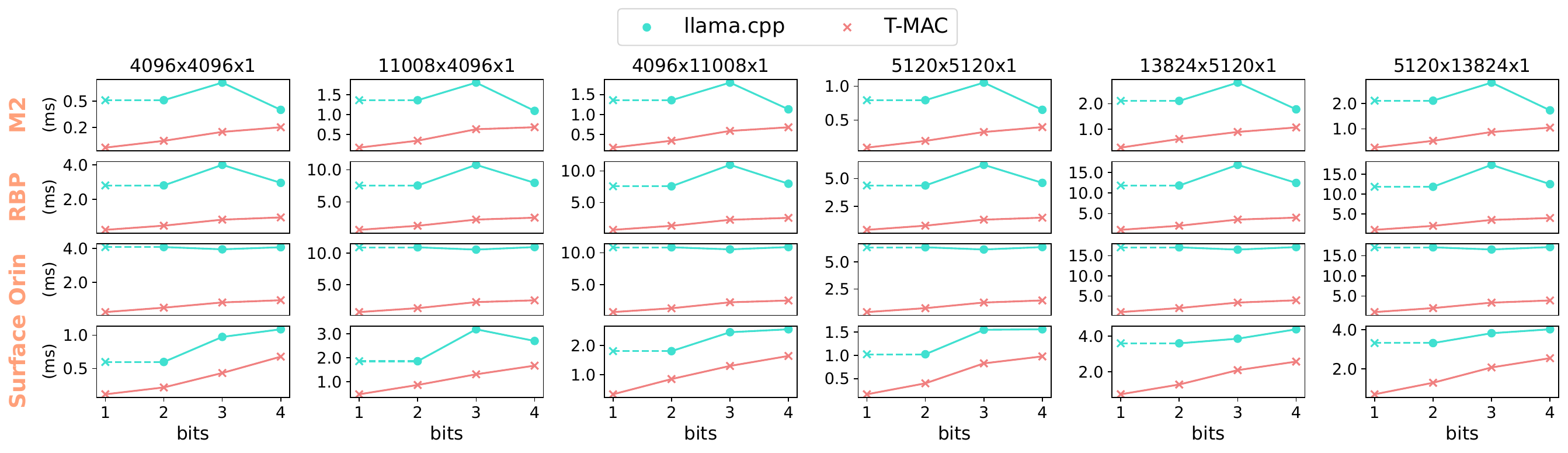}
    \vspace{-2.0em}
    \caption{Single-Threaded mpGEMV}
  \end{subfigure}
  \begin{subfigure}[b]{\linewidth}
    \includegraphics[width=\linewidth]{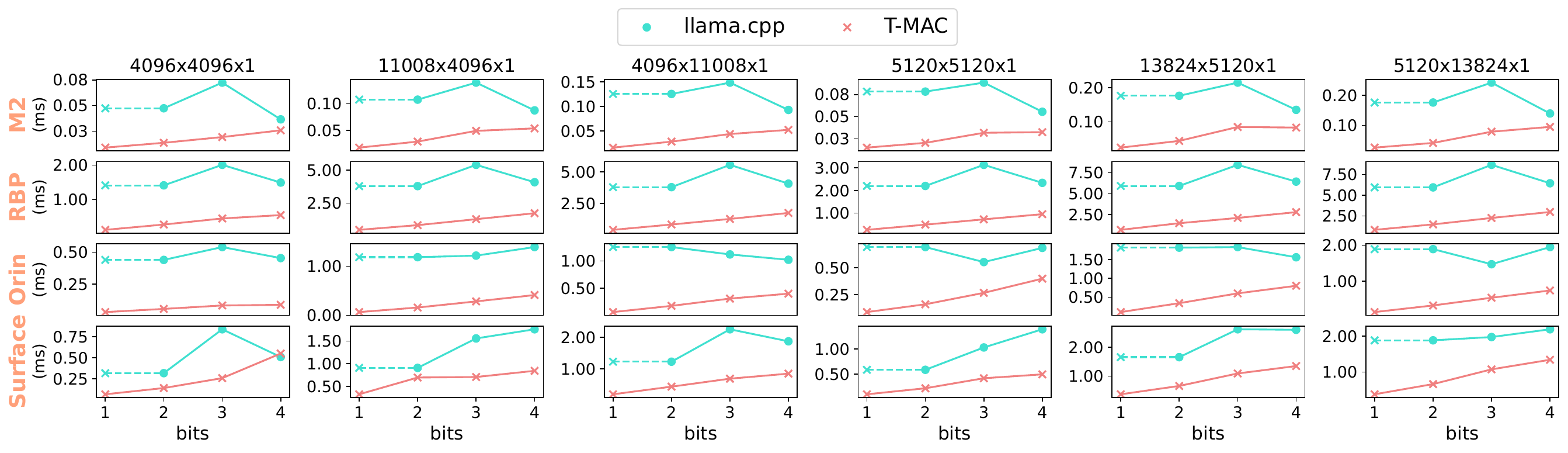}
    \vspace{-2.0em}
    \caption{Multi-Threaded mpGEMV}
  \end{subfigure}
  \caption{mpGEMV performance benchmark at 1/2/3/4 bits with single-thread and multi-thread. Matrix shapes are from Llama-2-7b and Llama-2-13b. The 1-bit kernel performance of llama.cpp is deduced from its 2-bit kernel and marked with dashed lines.}
  \label{fig:GEMV}
\end{figure*}

\paragraph{Hardware devices} As shown in Table~\ref{tab:spec}, we evaluate \sys across four distinct edge devices. These devices range from high-performance ones like M2-Ultra to less powerful ones like Raspberry Pi. The CPUs tested encompass Intel Core, Apple Silicon, and Cortex series. The operating systems include OSX, Linux, and Windows. This evaluation guarantees \sys's cross-platform compatibility and consistent performance across different instruction sets and various edge deployment scenarios.

\paragraph{Kernels and models}
To evaluate the performance of \sys{}, we conduct extensive benchmarks using real-word low-bit LLMs and scenarios.
For the kernel performance benchmark, we select matrix shapes derived from the Llama-2-7B and Llama-2-13B models, ensuring our evaluation reflects the practical demands.
To conduct an end-to-end throughput test, we employed actual quantized models to demonstrate the practical efficacy of \sys{} across different bit-width configurations.
Specifically, we employ 4-bit,3-bit,2-bit and 1-bit quantized Llama models, and also 1-bit and 1.58bit BitNet models that are trained from scratch. \revision{Ternary weights in 1.58bit BitNet are interpreted as 2-bit and decomposed into two 1-bit matrices.}
The 4-bit Llama models are from GPTQ~\cite{frantar2022gptq}.
The 3-bit and 2-bit Llama models are from BitDistiller~\cite{du2024bitdistiller}.
The 1-bit Llama models are from OneBit~\cite{xu2024onebit}.
    
\paragraph{Baselines.}
We compared the performance of \sys{} with \textit{llama.cpp} \revision{(version b2794, realesed on May 2024)}, a state-of-the-art implementation for LLM deployment on edge devices.
We chose \textit{llama.cpp} as the baseline for several compelling reasons. Firstly, \textit{llama.cpp} represents the cutting-edge in LLM deployment on edge devices, featuring highly optimized kernel implementations tailored to each hardware platform. Its versatility and robust performance make it an ideal benchmark for assessing the efficacy of new methodologies. Additionally, \textit{llama.cpp} is implemented in plain C/C++ without any dependencies, ensuring maximum compatibility and efficiency across diverse hardware configurations. For kernel performance benchmarks, we utilized the optimized kernels provided by \textit{llama.cpp} as the baselines on the respective hardware devices. In our end-to-end throughput evaluations, we integrate the LUT-based kernels from \sys{} to \textit{llama.cpp} and compare it with original \textit{llama.cpp}.

\revision{We also compared the performance of \sys{} with \textit{llama.cpp (BLAS)}. \textit{llama.cpp} uses \textit{Accelerate} on M2-Ultra and \textit{OpenBLAS} on the other platforms. \textit{llama.cpp (BLAS)} is slower for mpGEMV but faster for mpGEMM compared to \textit{llama.cpp}’s highly optimized mixed-precision implementation. Therefore, \sys{} is measured against \textit{BLAS} only for mpGEMM.}

\paragraph{Measurement approach.} We perform both \textit{kernel-level} and \textit{model-level} measurement. To obtain precise and consistent kernel-level latency on CPU, we first perform a warmup of 10 iterations, followed by 100 runs to calculate an average. The warmup on M2-Ultra differs slightly from the others, requiring at least 1 second to maximize performance. To perform model-level latency, we integrate \sys{} into llama.cpp. We repeatedly generate 64 tokens for 20 iterations to evaluate token generation throughput.

\begin{figure*}[t]
  \centering
  \begin{subfigure}[b]{\linewidth}
    \includegraphics[width=\linewidth]{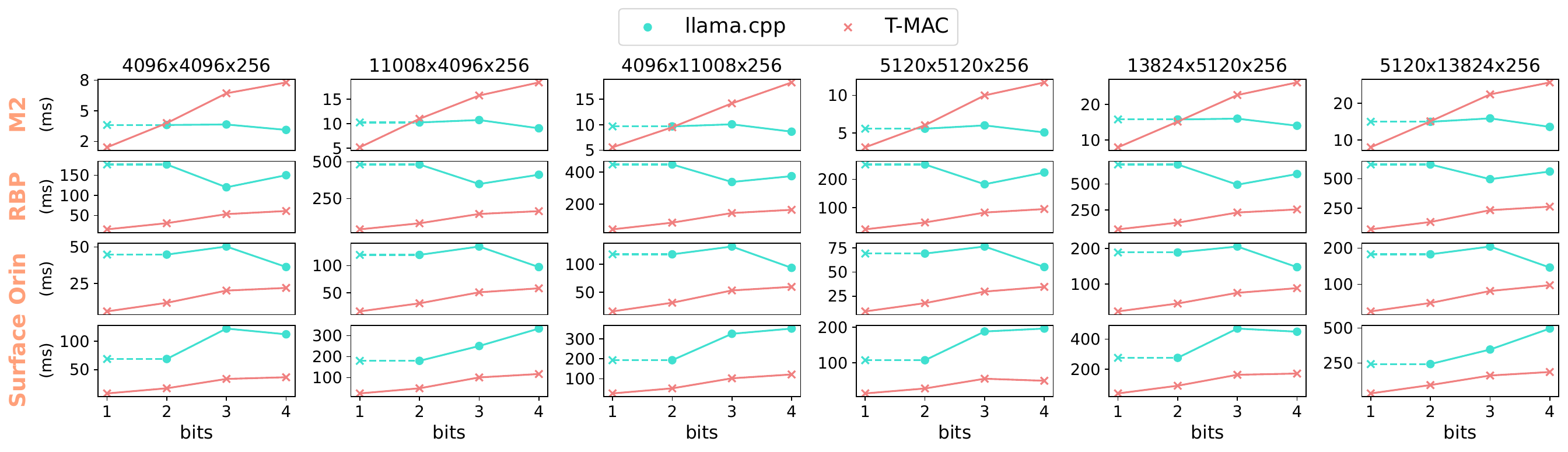}
  \end{subfigure}
  \vspace{-2.0em}
  \caption{mpGEMM performance benchmark at 1/2/3/4 bits with multi-thread. Matrix shapes are from Llama-2-7b and Llama-2-13b with an input sequence length of 256. The 1-bit kernel performance of llama.cpp is deduced from its 2-bit kernel and marked with dashed lines.}
  \label{fig:gemm}
\end{figure*}

\subsection{mpGEMV/mpGEMM Performance Benchmark}

We evaluate the kernels in Llama-2-7B/13B across all four devices. As illustrated in Figure~\ref{fig:GEMV} with the bits decrease from 4-bit to 2-bit, llama.cpp fails to gain any additional speedup, and even experiences a 15\% slowdown at 3-bit compared to 4-bit due to decoding overhead. Therefore, we can infer that the 1-bit llama.cpp performance would be similar to 2-bit, even though llama.cpp does not provide a 1-bit implementation. In contrast, \sys{} achieves linear speedup with bit reduction. For single-threaded GEMV, \sys achieves maximum speedups of 11.2x, 5.8x, 4.7x, and 3.1x respectively for 1/2/3/4 bits.
For multi-threaded mpGEMV, \sys's performance is primarily constrained by memory bandwidth, but \sys{} can still achieve significant speedup due to efficient memory access. For example, with 2-bit, \sys achieves 4.0x, 4.0x, 5.31 and 2.5x on all four devices respectively.

We evaluate mpGEMM with a sequence length of 256 for multi-threading in Figure~\ref{fig:gemm}. \textit{llama.cpp} uses BLAS for mpGEMM. \sys{} still achieves significant speedup on RBP, Orin, and Surface for 2-bit with maximum speedups of 4.0x, 5.3x, and 5.3x respectively. M2-Ultra is an exception, as the Apple Silicon CPUs are equipped with a powerful AMX coprocessor to handle GEMM operations. However, \sys{} still achieves a maximum 2.0$\times$ speedup  for 1-bit in this case.

\revision{\sys{} demonstrates a significant advantage at 3-bit precision. This can be attributed to the inefficiencies in current techniques for handling 3-bit weights. Weights decoding is typically executed using SHIFT and AND instructions, which require weights being aligned with the width of bytes. Since 8 is indivisible by 3, this decoding process is notably inefficient. \textit{llama.cpp} attempts to optimize it by separately packing 2 bits and the remaining 1 bit, but it still results in significant overhead. In contrast, \sys{} avoids this problem by individually computing the results of each bit.}

\subsection{End-to-End Inference Throughput}

\begin{figure}[t]
  \centering
  \includegraphics[width=\linewidth]{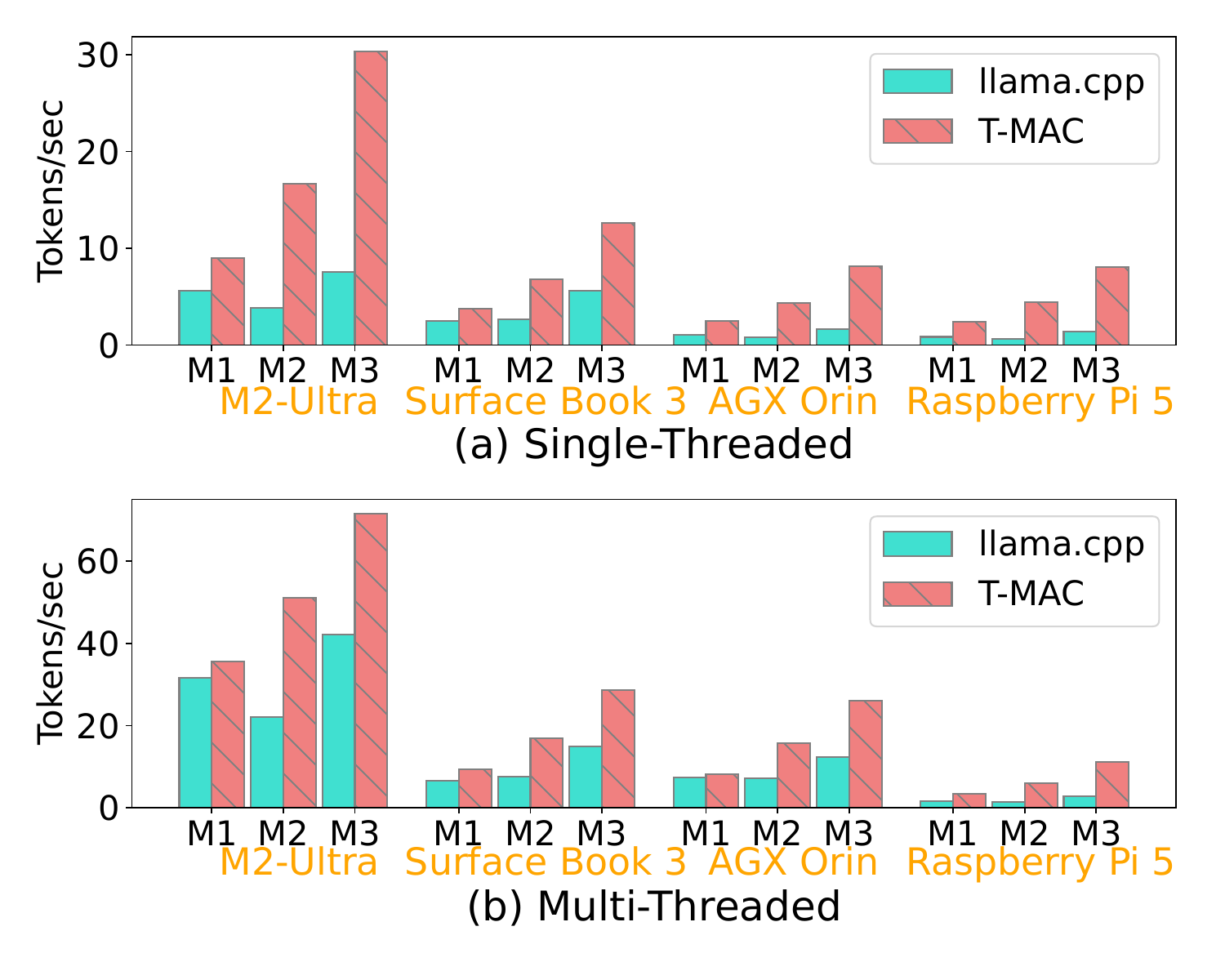}
  \vspace{-2.0em}
  \caption{End-to-end token generation throughput by integrating \sys{} kernels into llama.cpp. M1, M2 and M3 stands for Llama-2-7B-4bit, Llama-2-7B-2bit and BitNet-3B.}
  \label{fig:e2e}
\end{figure}

After integrating into llama.cpp, we compare the end-to-end token generation throughput of llama.cpp with \sys{}. We employ 2-bit to execute BitNet. As depicted in Figure~\ref{fig:e2e}, under single-threading on Raspberry Pi 5, \sys{} achieves speedups of 2.8x, 6.7x, and 5.8x for the three models respectively. Under multi-threading, due to memory constraints and operators other than mpGEMV/mpGEMM, the speedup is less pronounced. However, we still observe speedups of 1.1x, 2.3x, and 1.7x on M2-Ultra.
\sys{} can reach a peak of 71 tokens/sec on the powerful M2-Ultra and 11 tokens/sec on the less powerful Raspberry Pi 5, indicating promising real-world edge deployment.

\subsection{Power and Energy Consumption}

In addition to computational efficiency, energy efficiency is equally critical, especially for edge devices that rely on battery power.
To evaluate the power and energy consumption of \sys{} relative to \textit{llama.cpp}, we conducted experiments using multi-threaded implementations on the M2 Ultra device. We selected three models for our analysis: Llama-2-7B-4bit, Llama-2-7B-2bit, and BitNet-3B.
The power usage is measured using \textit{powermetrics} on OSX, which can record the average power usage over a specified sample interval. We set this interval to 500 milliseconds, continuously generate tokens for a minimum of 120 seconds, and compute the integral of power over time to determine the total energy consumption.

\begin{figure}[t]
    \centering
    \includegraphics[width=\linewidth]{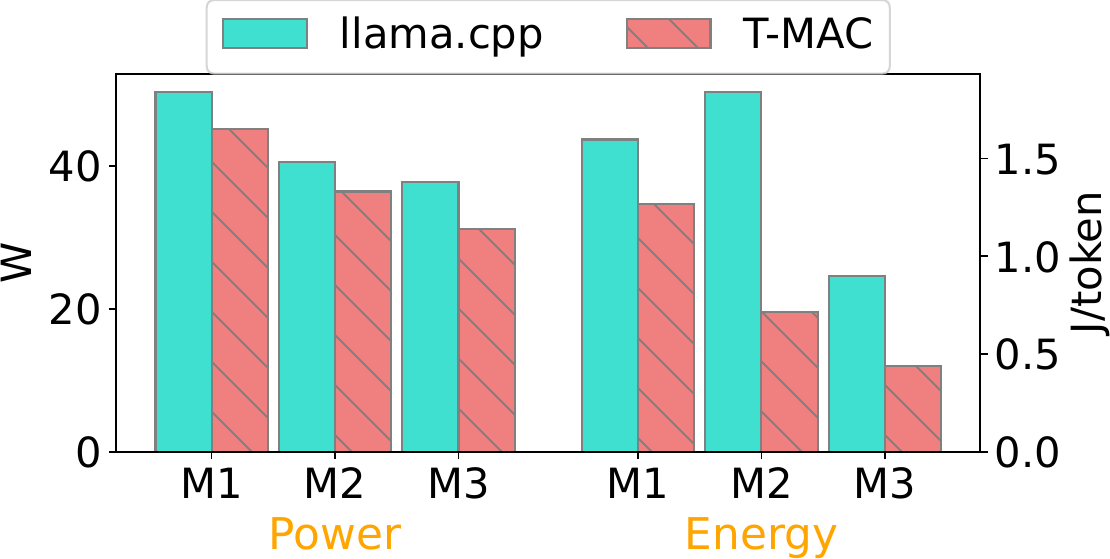}
    \caption{Power and energy consumption for multi-threaded inference on M2-Ultra. M1, M2 and M3 stands for Llama-2-7B-4bit, Llama-2-7B-2bit and BitNet-3B respectively.}
    \label{fig:e2e_power}
\end{figure}

The results, as depicted in Figure~\ref{fig:e2e_power}, indicate a significant reduction in power consumption when using the LUT-based kernels in \sys{}.
For the models Llama-2-7B-4bit, Llama-2-7B-2bit, and BitNet-3B, \sys{} demonstrates a power consumption reduction of 10.3\%, 10.3\%, and 17.3\%, respectively. These reductions in power consumption, combined with the latency gains offered by \sys{}, lead to a substantial decrease in total energy consumption. Specifically, \sys{} reduces energy consumption by 20.6\%, 61.2\%, and 51.3\% for each model, respectively.

\subsection{Optimization Breakdown}

To evaluate the effectiveness of the optimizations in \S\ref{sec:design}, we break down our optimization strategies. Most of these optimizations yield greater benefits for single-threading, but tiling requires multi-threading to be effective, hence this evaluation is conducted using multi-threading. While most optimizations are orthogonal, some are dependent on others, for instance, the permutation requires tiling to be completed first. Consequently, we begin with a basic implementation, \textit{TM-base}, and progressively apply optimizations. 

\begin{figure}[t]
    \centering
    \includegraphics[width=\linewidth]{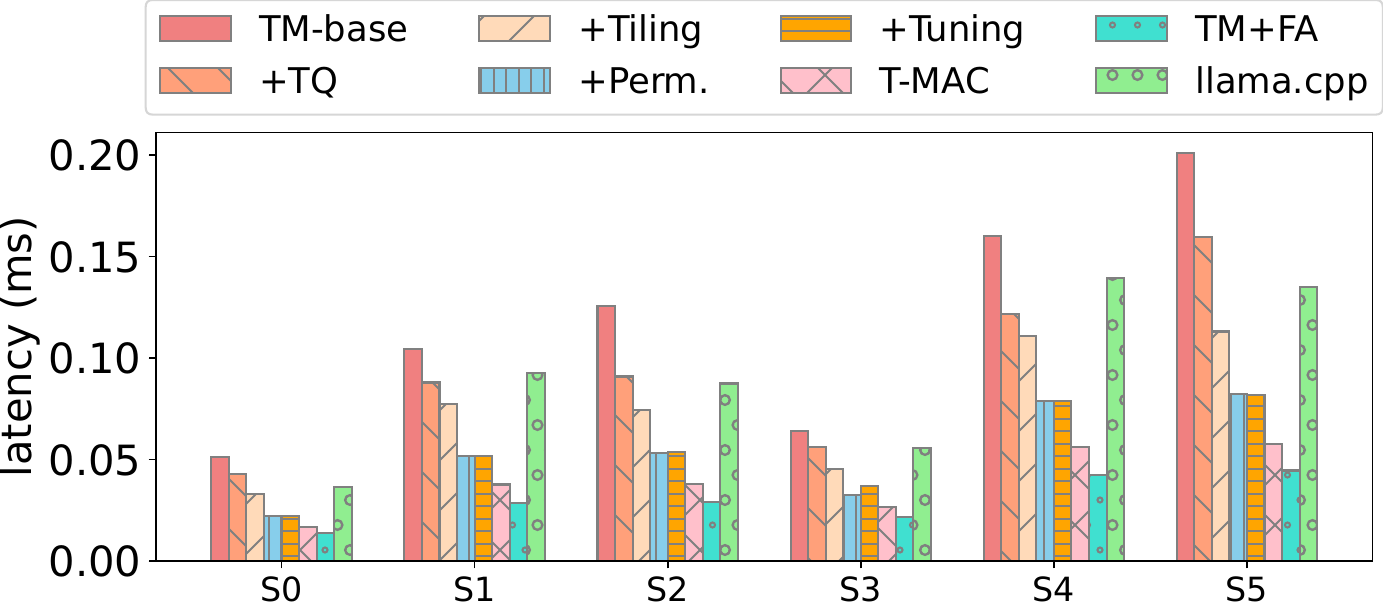}
    \caption{Multi-threaded performance of Llama-2-7B/13B GEMV kernels on M2-Ultra by applying \sys{} optimizations step-by-step. S0-S5: different shapes as shown in Figure~\ref{fig:GEMV}. TM: \sys{}, TQ: Table Quantization, Perm.: Permutation, IL: Interleaving, FA: Fast Aggregation.}
    \label{fig:ablation}
    \vspace{-2.0em}
\end{figure}

The \textit{TM-base} version utilizes hardware intrinsics to speed up table lookup, but does not implement any memory access optimization. As depicted in Figure~\ref{fig:ablation}, its performance is at most 17\% slower compared to the llama.cpp baseline. After implementing \textit{table quantization}, the performance becomes competitive with llama.cpp. The \textit{tiling} optimization further yields a maximum speedup of 1.45x. By rearranging the data layout into contiguous memory for each tiling, \textit{permutation} contributes an additional 1.39x speedup. \textit{Tuning} does not appear to be very effective in the figure, as the default tiling configurations already align well with M2-Ultra registers and caches, but for different devices, tuning should assist in finding a better configuration. 

Upon applying \textit{weights interleaving}, we obtain \sys{}. Interleaving eliminates most of the unpacking overhead, achieving a 1.42x speedup. The aggressive \textit{fast aggregation} can make \sys{} up to 1.29x faster, but it could lead to non-negligible accuracy loss, so we offer it as an optional optimization.

\subsection{Error Analysis} \label{error_analysis}

There are two sources of error compared to conventional mpGEMM implementation: (a) \textit{table quantization}, which is an algorithmic approximation included in our method, and (b) \textit{fast aggregation}, whose error is introduced during the instruction execution within the fixed CPU architecture. We evaluate the impact of these two error sources at both kernel-level and model-level. 

\paragraph{Kernel-level Evaluation.} We use the unquantized\\$W_{FP16}A_{FP16}$ GEMV as the benchmark. The weights and activation of the GEMV are randomly generated FP16 values following a \textit{Gaussian Distribution}, which are then quantized to 4-bit for execution by llama.cpp and \sys. The Normalized Mean Squared Error (NMSE) is then computed between the ground truth and the mpGEMV outputs. As shown in Table.~\ref{tab:nmse}, the NMSE difference between llama.cpp and \sys{} is negligible, indicating that the table quantization error is minimal. However, after applying \textit{fast aggregation}, the NMSE increases to 2.5$\times$.

\begin{table}[t]
    \centering
    \begin{tabular}{c c c c}
    \toprule
    \textbf{MxKxN} & \textbf{llama.cpp} & \textbf{\sys{}} & \textbf{\sys{} (+FA)} \\
    \midrule
    4096x4096x1 & 3.33e-03 & 3.35e-03 & 8.09e-03 \\
    11008x4096x1 & 3.44e-03 & 3.46e-03 & 8.27e-03 \\
    4096x11008x1 & 4.13e-03 & 4.15e-03 & 8.45e-03 \\
    \bottomrule
    \end{tabular}
    \caption{NMSE error relative to un-quantized ($W_{FP16}A_{FP16}$) GEMV kernel.}
    \label{tab:nmse}
    \vspace{-1.5em}
\end{table}

\begin{table}[t]
    \centering
    \resizebox{\linewidth}{!}{
    \begin{tabular}{lccccc}
    \toprule
    \multirow{2}*{\textbf{Framework}} & \textbf{Throughput} & \textbf{WikiText2} & \textbf{lambada\_openai} & \textbf{WinoGrande} \\
     & Tokens/sec $\uparrow$ & PPL $\downarrow$ & PPL $\downarrow$ & Acc. $\uparrow$ \\
    \midrule
    Un-quantized & 3.79 & 5.80 & 12.65 & 71.0 \\
    llama.cpp & 5.65 & 5.96 & 12.95 & 70.8 \\
    \sys{} & 7.34 & 5.96 & 12.95 & 70.8 \\
    \sys{} (+FA) & 8.97 & 6.38 & 13.99 & 67.8 \\
    \bottomrule
    \end{tabular}
    }
    \caption{End-to-end throughput and model quality of Llama-2-7B-4bit on M2-Ultra with single-thread. \sys{} improves throughput by 1.3$\times$ compared to llama.cpp with the same model quality. Fast Aggregation (FA) can further improve the throughput gain to 1.6$\times$, but the model quality will drop because of the numerical error of current CPU instructions.}
    \label{tbl:model-acc}
    \vspace{-2.0em}
\end{table}

\paragraph{Model-level Evaluation.} To examine the impact of these errors on real-world models, we chose Llama-2-7B for testing. The models are the GGUF model converted from official Llama-2-7B weights for the un-quantized ground truth and the original llama-2-7b.Q4\_0.gguf model~\cite{gguf-models} released with llama.cpp for mpGEMM. After integrating \sys{} into llama.cpp, we conduct the evaluation through the \textit{perplexity}~\cite{llamacpp-perplexity} tool provided by llama.cpp. The evaluation is performed on three different tasks: \textit{WikiText-2}~\cite{merity2016pointer} and \textit{lambada\_openai}~\cite{paperno-etal-2016-lambada,radford2019language} for perplexity (the lower the better), and \textit{WinoGrande}~\cite{ai2:winogrande} for question answering accuracy (the higher the better. As shown in Table~\ref{tbl:model-acc}, on all of the three tasks, \sys{} delivers the same results compared to llama.cpp, suggesting that the error introduced by \sys{} is negligible for real-world models. After toggling on the \textit{fast aggregation}, the perplexity increases by 0.4 and 1.0 respectively and the accuracy drops by 0.3\%.

In summary, \sys\ introduces negligible error to model inference while offering significant speedup. The \textit{fast aggregation} can further enhance performance, but at the cost of model quality. We offer this as an option for users in scenarios that prioritize real-time performance and are less sensitive to accuracy. Without \textit{fast aggregation}, \sys can still achieve substantial gain according to Figure~\ref{fig:ablation}. In the future, we anticipate the error introduced by fast aggregation can be mitigated with straightforward optimizations of the CPU micro-architecture.

\subsection{Compared with GPU\revision{/NPU}}

GPUs are widely used in LLM deployments. We compare \sys{} on CPU with llama.cpp on GPU to illustrate the efficiency of \sys{}.
llama.cpp is the state-of-the-art LLM inference framework for both CPU and GPU on edge devices.


\begin{figure}[t]
    \centering
    \includegraphics[width=\linewidth]{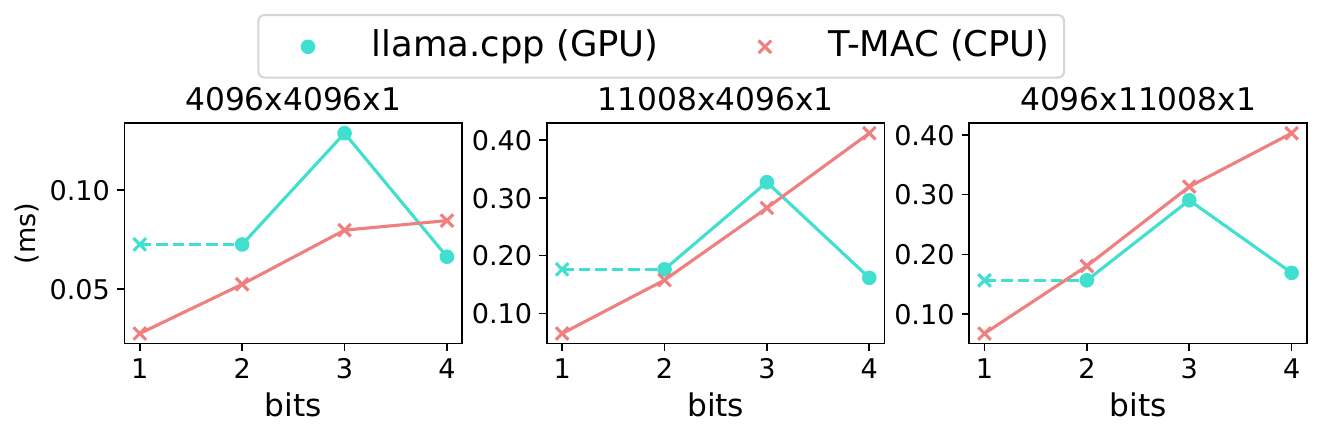}
    \caption{mpGEMV kernels performance of \sys{} (CPU) and llama.cpp (GPU) on NVIDIA Jetson AGX Orin.}
    \label{fig:gemv_cuda}
\end{figure}

Figure~\ref{fig:gemv_cuda} shows the mpGEMV kernel performance comparsion of \sys{} (CPU) and llama.cpp (GPU) on NVIDIA Jetson AGX Orin, a platform with ARM CPU and NVIDIA CUDA GPU. The kernel configurations are all from Llama-2-7B. \sys{} significantly outperforms GPU on W1A16 on all cases, while achieves comparable performance on W2A16 and W3A16. Although GPU performs better on higher bits and larger shape due to its powerful parallel computing capacity, this evaluation still shows huge potential of CPU-based LLM deployments on edge devices.

\begin{table}[t]
    \centering
    \resizebox{0.82\linewidth}{!}{
    \begin{tabular}{rccc}
    \toprule
    \multirow{2}*{\textbf{Framework}} & \textbf{Throughput} & \textbf{Power} & \textbf{Energy} \\
     & Tokens/sec & W & J/token \\
    \midrule
    llama.cpp (CPU) & 7.08 & 15.0 & 2.12 \\
    llama.cpp (GPU) & \textbf{20.03} & 30.8 & 1.54 \\
    \sys{} (CPU) & 15.62 & \textbf{10.4} & \textbf{0.66} \\
    \bottomrule
    \end{tabular}
    }
    \caption{Llama-2-7B-2bit end-to-end inference throughput, power and energy comparisons on NVIDIA Jetson AGX Orin.}
    \vspace{-5mm}
    \label{tab:orin-gpu}
\end{table}

\newcommand{\minitab}[2][l]{\begin{tabular}{#1}#2\end{tabular}}

\begin{table}[t]
\centering
\resizebox{\linewidth}{!}{
\begin{tabular}{c c c c c}
\toprule
\textbf{Device} & \textbf{Total CPU Cores} & \textbf{Used Cores} & \textbf{GPU} & \textbf{NPU} \\
\midrule
Surface  & \multirow{2}{*}{12 Oryon, 3.8 GHz} & \multirow{2}{*}{4} & Adreno X1-85 & Hexagon \\
Laptop 7 &                                    &                    & 4.6 TFLOPS   & 45 TOPS \\
\midrule
\multirow{3}{*}{OnePlus 12} & 1 Cortex-X4, 3.2 GHz   & 1 & \multirow{3}{*}{\minitab[c]{Adreno 750\\4.6 TFLOPS}} & \multirow{3}{*}{\minitab[c]{Hexagon\\15 TOPS}} \\
                            & 5 Cortex-A720, 3.0 GHz & 3 \\
                            & 2 Cortex-A520, 2.3 GHz & 0 \\
\midrule
Jetson  & \multirow{2}{*}{8 Cortex A78AE, 2.0 GHz} & \multirow{2}{*}{6} & Ampere GA10B & \multirow{2}{*}{-} \\
Orin NX &                                          &                    & 50 TOPS      \\
\bottomrule
\end{tabular}
}
\caption{\revision{Detailed CPU/GPU/NPU specifications of the tested platforms. "Used Cores" refers to the number of CPU cores we used to fully utilize the memory bandwidth.}}
\vspace{-2.0em}
\label{tab:spec-ext}
\end{table}

\begin{table*}[t]
    \centering
    \begin{tabular}{c c c c c c}
    \toprule
    \textbf{Device} & \textbf{Model} & \textbf{\sys{} (CPU)} & \textbf{llama.cpp (CPU)} & \textbf{llama.cpp (GPU)} & \textbf{NPU} \\
    \midrule
    \multirow{2}{*}{Surface Laptop 7} & Llama-2-7B-4bit & \textbf{21.63} & 10.64 & \multirow{2}{*}{-} & 10.40  \\
                                      & Llama-2-7B-2bit & \textbf{31.83} & 9.39  &                    & 10.40* \\
    \midrule
    \multirow{2}{*}{OnePlus 12} & Llama-2-7B-4bit & 10.19          & 8.24 & 1.60 & \textbf{11.30}  \\
                                & Llama-2-7B-2bit & \textbf{16.62} & 6.95 & 1.72 & 11.30*          \\
    \midrule
    \multirow{2}{*}{Jetson Orin NX} & Llama-2-7B-4bit & 7.53           & 3.97 & \textbf{14.76} & \multirow{2}{*}{-} \\
                                    & Llama-2-7B-2bit & \textbf{11.41} & 3.20 & 7.94           \\
    \bottomrule
    \end{tabular}
    \caption{\revision{Token generation tokens/s of \sys{} vs GPU/NPU for Llama-2-7B-4bit/2bit on three devices. The 2-bit performance of NPUs is deduced from 4-bit and marked with "*".}}
    \label{tab:cpu-gpu-npu}
    \vspace{-1.0em}
\end{table*}

Table~\ref{tab:orin-gpu} shows the end-to-end comparison of the Llama-2-7B-2bit model on NVIDIA Jetson AGX Orin. Without \sys{}, CPU only performs better than GPU in power, however, the energy consumption is still worse than GPU due to lower throughput. Compared to llama.cpp on CPU, \sys{} not only improves the throughput to 2.2$\times$, but also reduces the power to 69$\%$, resulting in 3.2$\times$ energy efficiency. Compared to llama.cpp on GPU, although \sys{} only achieves 78$\%$ throughput, \sys{} only needs 34$\%$ power, resulting in 2.3$\times$ energy efficiency. Note that Figure~\ref{fig:gemv_cuda} shows \sys{} outperforms the GPU on the mpGEMV kernels. The reason why the throughput of \sys{} is still lower than that of GPU is due to the performance of kernels except mpGEMVs in llama.cpp on CPU.

\revision{
In addition to its power efficiency, \sys{} also demonstrates superior performance over GPU/NPUs across widely used platforms. We further evaluate \sys{} on three devices: Surface Laptop 7, OnePlus 12 and Jetson Orin NX. The full specifications of equipped CPU/GPU/NPUs are detailed in Table~\ref{tab:spec-ext}. We utilize the minimum number of CPU cores that can fully leverage the memory bandwidth and nearly achieve optimal performance. GPU evaluations use the llama.cpp CUDA backend for the NVIDIA GPU and OpenCL backend for the Qualcomm GPU. The performance of NPUs are sourced from official data released by Qualcomm via Qualcomm AI Hub~\cite{qualcomm-ai-hub}.

With T-MAC, CPUs can achieve much higher computation throughput and fully exploit the memory bandwidth. GPU/NPUs, which are also memory-bound for GEMV during token generation, share unified memory with CPUs on most edge devices. As demonstrated in Table~\ref{tab:cpu-gpu-npu}, \sys{} achieves significant speedup for Llama-2-7B-2bit on all three devices. Specifically, \sys{} achieves 3$\times$ speedup on Surface Laptop 7 over the NPU with only 4 out of the total 12 CPU cores, 1.5$\times$ speedup over the NPU on OnePlus 12, and 1.4$\times$ speedup over the Ampere GPU on the Jetson Orin NX. Even for Llama-2-7B-4bit, \sys{} maintains 2.1$\times$ speedup on Surface Laptop 7. Notably, on OnePlus 12, \sys{} demonstrates a substantial speedup of 6.4$\times$ and 9.7$\times$ over the Adreno GPU for 4-bit and 2-bit respectively.

In summary, \sys{}, leveraging widely available CPUs, delivers a notable performance advantage over GPUs, and even NPUs specially designed for AI workloads. This makes \sys{} a practical solution for LLM deployment on edge devices.
}

\section{Related Works}
\paragraph{LLM Quantization Algorithm}
LLM quantization has emerged as a crucial technique for the efficient deployment of LLMs in resource-constrained environments.
A segment of the research has been dedicated to the dual quantization of both weights and activations.
\textit{LLM.int8()}~\cite{dettmers2022gpt3.int8} isolates outlier feature dimensions to 16-bit computations while processing majority dimensions in efficient 8-bit computations.
\textit{SmoothQuant}~\cite{xiao2023smoothquant} migrates the quantization difficulty from activations to weights with a mathematically equivalent transformation to enable an INT8 quantization of both weights and activations.
Advancements in the field have led to a refined focus on the singular quantization of model weights, as weight storage accounts for the majority of memory footprint.
Specific algorithms such as \textit{GPTQ}~\cite{frantar2022gptq} and \textit{AWQ}~\cite{lin2023awq} have demonstrated the feasibility of quantizing LLMs to just 4 bits using post-training quantization techniques.
Furthermore, \textit{BitDistiller}~\cite{du2024bitdistiller} pushes the boundary to 2 bits by leveraging quantization-aware training (QAT) and self-distillation.
Meanwhile, \textit{BitNet}~\cite{wang2023bitnet} takes an even more ambitious route by training 1-bit LLMs from scratch.

\paragraph{LLM Inference System}
The significance of LLMs has spurred the development of various LLM inference systems, tailored to different platforms and optimized for specific goals.
\textit{vLLM}~\cite{vllm} is a high-throughput and memory-efficient inference engine designed for LLMs, which excels in large batch processing.
\textit{llama.cpp}~\cite{llama.cpp} stands out with its plain C/C++ implementation, free of external dependencies, which delivers superior performance on edge computing devices.
\textit{TensorRT-LLM}~\cite{trtllm} incorporates a suite of state-of-the-art optimizations specifically for NVIDIA GPUs.
\textit{Intel Neural Compressor}~\cite{intel_neural_processor} provides an open-source Python library for model compression techniques, tailored to enhance the efficiency of LLMs within the Intel ecosystem.
All of these inference systems share a crucial capability for supporting low-bit LLMs, which not only minimizes memory usage but also enhances computational efficiency, thereby broadening the accessibility of LLMs for diverse applications.
To complement the landscape of end-to-end LLM inference systems, there are also efforts concentrated on developing highly efficient computational kernels tailored for low-bit LLMs~\cite{frantar2024marlin,bitblas}.

\section{Conclusion}
\sys{} transforms the data-type-centric multiplication to bit-wise table lookup, and provide a unified and scalable solution for the increasingly popular mpGEMM. On the pervasively available CPUs of edge devices, \sys{} kernels achieve up to 6.6$\times$ speedup compared to llama.cpp, which makes the CPU inference speed comparable or even higher than the GPU on the same device. \sys{} thus provides a practical solution to deploy LLMs on edge devices without relying on GPU, even on a Raspberry Pi. \sys{} also opens up the broad opportunity for novel LLM hardware accelerator design based on LUT, as LUT is much more efficient in hardware implementation than multiplications.

\bibliographystyle{plain}
\bibliography{references}

\end{document}